\documentclass[11pt]{article}
\usepackage{amssymb,amsmath}
\usepackage{ascmac}
\usepackage[dvipdfmx]{hyperref}
\usepackage[dvips]{graphicx,psfrag}
\usepackage{epsfig}
\usepackage{array}
\usepackage{lscape}
\usepackage{accents}
\usepackage{color}
\usepackage{braket}
\usepackage{bm}
\usepackage{ulem}

\numberwithin{equation}{section}

\makeatletter
\def\alt{\mathrel{\mathpalette\gl@align<}}
\def\agt{\mathrel{\mathpalette\gl@align>}}
\def\gl@align#1#2{\lower.6ex\vbox{\baselineskip\z@skip\lineskip\z@
\ialign{$\m@th#1\hfil##\hfil$\crcr#2\crcr\sim\crcr}}}
\makeatother

\newcommand{\doi}[1]{\href{https://doi.org/#1}{doi:#1}}

\newcommand{\mubar}{\overline{\text{Mu}}}
\newcommand{\dash}{\hspace{1pt}--\hspace{1pt}}

\begin{document}

\begin{flushright}
{\large
April, 2022
}
\end{flushright}
\vspace*{5mm}

\begin{center}

{\bf
\LARGE
Transverse positron polarization in the polarized $\mu^+$ decay  
 related with the muonium-to-antimuonium transition
} 

\vspace{1.0cm}

{\Large 
Takeshi Fukuyama$\,{}^{a}$,
Yukihiro Mimura$\,{}^{b}$
and
Yuichi Uesaka$\,{}^{c}$
}

\vspace{8mm}

{\large
${}^a${\it 
Research Center for Nuclear Physics (RCNP),
Osaka University, \\Ibaraki, Osaka, 567-0047, Japan
}\\
\vspace{3mm}
${}^{b}${\it
Department of Physical Sciences, College of Science and Engineering, \\
Ritsumeikan University, Shiga 525-8577, Japan
}\\
\vspace{3mm}
${}^{c}${\it
Faculty of Science and Engineering, Kyushu Sangyo University, \\
2-3-1 Matsukadai, Higashi-ku, Fukuoka 813-8503, Japan
}\\
}

\vspace{1.2cm}

{\Large
{\bf Abstract}}\end{center}
\baselineskip 18pt
{\large
The constructions of the new high-intensity muon beamlines are progressing
in facilities around the world, 
and new physics searches related to the muons are expected.
The facilities can observe 
the transverse positron polarization of the polarized $\mu^+$ decay
to test the standard model.
The transition of muonium into antimuonium (Mu-to-$\mubar$ transition), 
which is one of the interesting possibilities in the models beyond the standard model,
can be also tested.
The near-future observation of the transition gives us a great impact
since it indicates
that there is an approximate discrete symmetry in the lepton sector.
If the Mu-to-$\mubar$ transition operator is generated, a new muon decay
operator can exist and it may interfere with the standard model muon decay operator
to induce the corrections to the transverse positron polarization
in the $\mu^+$ decay.
In this paper,
we examine the possibility that
the Mu-to-$\mubar$ transition and the correction to the transverse positron polarization
are related, and
we show that those two are related
in the model of a neutral flavor gauge boson.
We also investigate the models to generate the Mu-to-$\mubar$ transition,
such as an inert $SU(2)_L$ doublet, a $SU(2)_L$ triplet for the type-II seesaw model, 
a dilepton gauge boson, and a left-right model.
The non-zero value of the transverse polarization for one of the two directions, $P_{\rm T_2}$,
violates the time-reversal invariance,
and the experimental constraint of the electron electric dipole moment can provide
a severe constraint on $P_{\rm T_2}$ depending on the model.
}


\thispagestyle{empty}
\newpage
\addtocounter{page}{-1}

\section{Introduction}

\fontsize{13pt}{18pt}\selectfont

The operating high-intensity muon beamlines 
at Japan Proton Accelerator Research Complex (J-PARC)
are being upgraded \cite{Kawamura:2018apy},
and 
muon fundamental properties, such as the anomalous magnetic moment ($g-2$)
and electric dipole moment (EDM),
will be accurately examined \cite{Abe:2019thb}.
The high-intensity muon beamlines are planned in the world \cite{Aiba:2021bxe}.
The facilities can produce 
the Muonium (a bound state of $\mu^+ e^-$),
and 
they will examine the muonium-to-antimuonium (Mu-to-$\mubar$) transition~\cite{Kawamura:2021lqk,Han:2021nod}, which is an interesting phenomenological possibility 
with the lepton flavor violation (LFV)~\cite{Pontecorvo:1957cp,Feinberg:1961zza,Lee:1977tib,Halprin:1982wm}.
By using the beamline, the transverse positron polarization in the polarized $\mu^+$ decay~\cite{Kinoshita:1957zza,Scheck:1977yg,Danneberg:2005xv,Fetscher:2021ldh}
will be also measured to find a clue of new physics \cite{Fukuyama:2019jiq}.

The LFV is one of the keys
of the new physics in the lepton sector
since it directly indicates that there is a new particle and interaction 
beyond the standard model (SM)
at the TeV scale.
The LFV processes, such as $\mu \to e\gamma$, $\mu \to 3e$ decays, and $\mu$\dash$e$ conversion in nuclei,
are not yet observed, 
and non-observation only gives severe bounds to the model parameters at present \cite{Adam:2013mnn,Bellgardt:1987du,SINDRUMII:2006dvw}.
We have to remark that
the absence of such $\Delta L_e$, $\Delta L_\mu = \pm 1$ processes 
does not necessarily mean that there is no new physics at the TeV-scale.
Even if those processes are absent,
there is still plenty of room left for new physics in the lepton sector.
Contrary to the quark sector,
 the lepton sector may have a high affinity with discrete symmetry;
e.g., the atmospheric neutrino mixing is nearly maximal.
Though it is surely important to search for the $\Delta L_e$, $\Delta L_\mu = \pm 1$ processes,
we need a close examination of the models beyond the SM
so as not to have a preconception that those processes are dominant
in the new physics with LFV.
Indeed, if there is an approximate discrete flavor symmetry,
the Mu-to-$\mubar$ transition as a $\Delta L_e$, $\Delta L_\mu = \pm 2$ process
can be important to find new physics in the lepton sector
while the $\Delta L_e$, $\Delta L_\mu = \pm 1$ processes are suppressed.
Since the
constraints from the $\Delta L_e$, $\Delta L_\mu = \pm 1$ processes
to obtain the Mu-to-$\mubar$ transition has been intensively investigated in Ref.\cite{Fukuyama:2021iyw},
we assume that 
the $\Delta L_e$, $\Delta L_\mu = \pm 1$ processes are absent by a discrete symmetry
in order to make the statements simple in this paper.

Let us suppose that the Mu-to-$\mubar$ transition rate is just below the current experimental bound.
We take notice of the existence of a new muon decay operator,
if (at least) one of the two muons and one of the two electrons are left-handed in the transition operator.
The coupling strength of the four-fermion operator 
is less than $O(10^{-3})$
in the unit of the Fermi coupling constant of the $(V-A)\times (V-A)$ muon decay in the SM
from the experimental result at the Paul Scherrer Institute (PSI) \cite{Willmann:1998gd}.
If the new effective operator for the muon decay is the type of $(S-P)\times (S+P)$,
the interference of the decay amplitudes can contribute to the 
transverse polarization of the $e^{\pm}$ from the polarized $\mu^{\pm}$ decay
in the primary order of the new physics.
Though the new coupling is bounded by the result of the Mu-to-$\mubar$ transition,
near-future experiments have the potential to observe 
the contributions from the new physics in the transverse positron polarization.
Actually, the high-intensity muon beam facility can examine both the Mu-to-$\mubar$ transition
and the transverse positron polarization in the polarized $\mu^+$ decay.

There are two independent transverse directions, and the positron polarizations are named
$P_{\rm T_1}$ and $P_{\rm T_2}$.
Let ${\bm{k}}_e$ be the momentum of the positron and ${\bm P}_\mu$ be the polarization vector 
(which specifies the degree and the direction of the polarization) of $\mu^+$ at rest.
The direction of $P_{\rm T_2}$ is defined to be that of ${\bm k}_e \times {\bm P}_\mu$.
A non-zero value of $P_{\rm T_2}$ violates the time-reversal invariance (namely, CP invariance),
and $P_{\rm T_2}$ is extremely tiny in the SM.
On the other hand,
 $P_{\rm T_1}$ is non-zero even in the SM,
 and its size is $\sim  m_e/m_\mu$ and becomes smaller for larger positron energy.
Observing the transverse polarizations gives us a useful test of
the fundamental interaction in the lepton sector.

In this paper, 
we examine the relation between the Mu-to-$\mubar$ transition 
and the transverse positron polarization in the polarized $\mu^+$ decay.
The four-lepton operators (without right-handed neutrinos) 
are generated by the tree-level exchange
of the inert $SU(2)_L$ doublet, $SU(2)_L$ triplet, 
dilepton gauge boson, and neutral flavor gauge boson.
We show that they are related
in the case of the neutral flavor gauge boson,
and the model can be tested in the near-future experiment.
The CP phases in the models
are severely bounded by the experimental constraint 
of the electron EDM (eEDM).
We discuss whether the eEDM can allow the non-zero $P_{\rm T_2}$
in the models.
We also describe the transverse positron polarization in the left-right model.

This paper is organized as follows:
In Section \ref{sec2}, we review the formulation of the muon decay
and give the formula for
the transverse polarization of the decayed $e^\pm$ 
in the polarized $\mu^\pm$ decay.
In Section \ref{sec3}, we review the expressions
of the Mu-to-$\mubar$ transition.
In Section \ref{sec4}, we describe the model of the neutral flavor gauge boson
and show
the relation between the
Mu-to-$\mubar$ transition and the transverse positron polarization
in the $\mu^+$ decay.
In Section \ref{sec5}, we consider the other models which can generate
the transverse positron polarization by the tree-level exchange of mediators.
Section \ref{sec6} is devoted to the conclusion.
In Appendix \ref{appendix:A}, we will give expressions of the Fierz transformation
of the muon decay operators for Majorana neutrinos.
In Appendix \ref{appendix:ref}, we comment on the physical background of the model discussed in Section \ref{sec4}.
In Appendix \ref{appendix:B}, 
we describe the constraints on the heavy-light neutrino mixing 
to evaluate the transverse positron polarization from the muon decay operators with 
right-handed neutrinos.

\section{Transverse polarization of the decayed $e^\pm$ in the $\mu^\pm$ decay}
\label{sec2}

In this section, we review the formalism for the transverse polarizations of the decayed $e^\pm$ in the polarized $\mu^\pm$ decay \cite{Kinoshita:1957zza,Scheck:1977yg,Fetscher:2021ldh}.
We follow the convention given in the review by the Particle Data Group \cite{Zyla:2020zbs}.
In general, we can write the four-fermion interaction for the $\mu \to e \nu_\mu \bar \nu_e$ decay 
in the Lagrangian \cite{Fetscher:1986uj} by 
\begin{align}
-\mathcal{L}_{\mu \to e \nu_\mu \bar \nu_e}&= \frac{4G_F}{\sqrt{2}}\sum_{\gamma=S,V,T}\sum_{\epsilon,m=R,L}g_{\epsilon m}^\gamma\left(\overline{e}_\epsilon\Gamma^\gamma\nu_e\right)\left(\overline{\nu}_{\mu}\Gamma_\gamma\mu_m\right)+{\rm H.c.},
\label{eq:general_4-Fermi_interaction}
\end{align}
where $G_F$ is the Fermi constant,
$\Gamma_S = {\bf 1}$, $\Gamma_V = \gamma_\mu$, and $\Gamma_T = \sigma_{\mu\nu}/\sqrt2$. 
For simplicity to describe, we call the operators by using the dimensionless couplings, $g_{\epsilon m}^\gamma$.
Since the two operators of $g_{LL}^T$ and $g_{RR}^T$ are identically zero, there are ten independent couplings.
The standard model corresponds to $g_{LL}^V=1$ and the other couplings being zero.
We note that 
the flavor violating ``wrong'' muon decay $\mu \to e \nu_e \bar\nu_\mu$ 
can interfere with the $\mu \to e \nu_\mu \bar \nu_e$ decay if the neutrinos are Majorana fermions.
We list the Fierz transformation of the operators for the Majorana neutrinos in Appendix \ref{appendix:A}.

When we neglect the radiative correction, the differential decay rate of $\mu^\pm$ is given by
\begin{align}
\frac{d^2\Gamma}{dxd\cos\theta} &= \frac{\bar G_F^2}{4\pi^3}m_\mu W_{e\mu}^4\sqrt{x^2-x_0^2}\left\{F_\mathrm{IS}(x)\pm P_\mu\cos\theta F_\mathrm{AS}(x)\right\}\left\{1+\hat{\zeta}\cdot\bm{P}_e\left(x,\theta\right)\right\}
\label{eq:decay_rate}
\end{align}
for the emitted $e^{\pm}$ with its spin parallel to the arbitrary direction $\hat{\zeta}$.
Here $P_\mu$ is the magnitude of the $\mu^\pm$ polarization vector $\bm{P}_\mu$, and $\theta$ is the angle between $\bm{P}_\mu$ and the $e^\pm$ momentum $\bm{k}_e$.
Defining the maximal $e^{\pm}$ energy,
\begin{align}
W_{e\mu}=\ &\frac{m_\mu^2+m_e^2}{2m_\mu},
\end{align}
we use the dimensionless variables,
\begin{align}
x& = \frac{E_e}{W_{e\mu}}, \hspace{3mm} x_0=\frac{m_e}{W_{e\mu}},
\end{align}
instead of the energy $E_e$ and the rest energy $E_0 = m_e$.
The allowed value of $x$ is between $x_0$ and 1.
We note that $\bar G_F$ in Eq.~\eqref{eq:decay_rate} includes the new physics effect 
$\bar G_F^2 = G_F^2 A/16$. See below for the parameter $A/16 \simeq 1$.
The muon decay constant determined by the muon lifetime is $\bar G_F$.
As we will note in the next section,
the quadratic corrections from the new muon decay couplings 
are bounded 
from the precision data relating to the universality of the weak couplings.

The polarization vector $\bm{P}_e$ of $e^\pm$ is defined 
by the differential decay rate in Eq.(\ref{eq:decay_rate}),
and the transverse components of $\bm{P}_e$ are defined by
\begin{equation}
P_{\rm T_1} \equiv \hat x_1 \cdot \bm{P}_e ,\qquad
P_{\rm T_1} \equiv \hat x_2 \cdot \bm{P}_e ,
\end{equation}
where the following
three unit vectors are defined by using the two specific directions $\bm{k}_e$ and $\bm{P}_\mu$:
\begin{align}
\hat{x}_3&= \frac{\bm{k}_e}{\left|\bm{k}_e\right|}, \quad \hat{x}_2=\frac{\bm{k}_e\times\bm{P}_\mu}{\left|\bm{k}_e\times\bm{P}_\mu\right|}, \quad \hat{x}_1=\hat{x}_2\times\hat{x}_3.
\end{align}
The transverse components can be written as
\begin{align}
P_\mathrm{T_1}\left(x,\theta\right)& 
=\frac{P_\mu\sin\theta F_\mathrm{T_1}(x)}{F_\mathrm{IS}(x)\pm P_\mu\cos\theta F_\mathrm{AS}(x)}, 
\label{eq:PT1}\\
P_\mathrm{T_2}\left(x,\theta\right) & 
=\frac{P_\mu\sin\theta F_\mathrm{T_2}(x)}{F_\mathrm{IS}(x)\pm P_\mu\cos\theta F_\mathrm{AS}(x)},
\end{align}
where the functions $F_{\rm IS}$, $F_{\rm AS}$, $F_{\rm T_1}$, and $F_{\rm T_2}$
depend on the coefficients $g_{\epsilon m}^\gamma$.
Instead of using the coefficients directly, it is practical to define the muon decay parameters 
for the spectrum and the transverse polarization
\cite{Michel:1949qe,Bouchiat:1957zz,Kinoshita:1957zz,Kinoshita:1957zza}.
For example, the function $F_{\rm T_2}$ is written as
\begin{equation}
F_\mathrm{T_2}(x)=\frac{1}{3}\sqrt{x^2-x_0^2}\left[3\frac{\alpha'}{A}(1-x)+2\frac{\beta'}{A}\sqrt{1-x_0^2}\right],
\end{equation}
where
\begin{align}
\alpha' &= 8\,\mathrm{Im}\left[g_{LR}^V\left(g_{RL}^{S*}+6g_{RL}^{T*}\right)-g_{RL}^V\left(g_{LR}^{S*}+6g_{LR}^{T*}\right)\right], \label{eq:alpha-prime} \\
\beta'& =4\,\mathrm{Im}\left[g_{RR}^Vg_{LL}^{S*}-g_{LL}^Vg_{RR}^{S*}\right].
\end{align}
It is important that the non-zero value of $P_\mathrm{T_2}$ indicates the CP violation in the interaction.
See Refs.\cite{Scheck:1977yg,Fetscher:2021ldh,Zyla:2020zbs} for the definition of the other CP conserving parameters 
($a,a',b,b',c,c',\alpha,\beta$; or their recombination, $\rho, \delta, \eta, \eta',  \xi, \xi', \xi^{\prime\prime},A$) and the functions, $F_{\rm IS}$, $F_{\rm AS}$, and $F_{\rm T_1}$.
Here, we write about the case when
there are only the following relevant operators for our purpose:
\begin{align}
-\mathcal{L}_{\mu \to e \nu_\mu \bar \nu_e}& = \frac{4G_F}{\sqrt{2}}\left[g_{LL}^V\left(\overline{e}\gamma^\alpha P_L\nu_e\right)\left(\overline{\nu}_{\mu}\gamma_\alpha P_L\mu\right)+g_{RR}^S\left(\overline{e}P_L\nu_e\right)\left(\overline{\nu}_{\mu} P_R\mu\right)\right. \nonumber\\
& \qquad \qquad \left.+g_{RR}^V\left(\overline{e}\gamma^\alpha P_R\nu_e\right)\left(\overline{\nu}_\mu(x)\gamma_\alpha P_R\mu\right)+g_{LL}^S\left(\overline{e}P_R\nu_e\right)\left(\overline{\nu}_{\mu} P_L\mu\right)\right]+ {\rm H.c.}
\label{eq:4-Fermi_interaction}
\end{align}
We obtain
\begin{align}
a &= a'=\alpha=\alpha'=c=c'=0, \\
b& = 4\left(\left|g_{LL}^V\right|^2+\left|g_{RR}^V\right|^2\right)+\left|g_{RR}^S\right|^2+\left|g_{LL}^S\right|^2, \\
b'& = -4\left(\left|g_{LL}^V\right|^2-\left|g_{RR}^V\right|^2\right)+\left|g_{RR}^S\right|^2-\left|g_{LL}^S\right|^2, \\
\beta &=4\ \mathrm{Re}\left[-g_{RR}^Vg_{LL}^{S*}-g_{LL}^Vg_{RR}^{S*}\right], \label{eq:beta} \\
\beta'&= 4\ \mathrm{Im}\left[g_{RR}^Vg_{LL}^{S*}-g_{LL}^Vg_{RR}^{S*}\right], \label{eq:beta-prime}
\end{align}
and the parameters for the muon decay spectrum (at the tree level) are given by
\begin{align}
\rho = \ &\delta=\frac{3}{4}, \hspace{3mm}  \eta= \eta'=0,  \hspace{3mm} \xi''=1, \\
\xi=\ &\xi'=-\frac{4b'}{A}, \\
A \equiv \ &a+4b+6c = 4b .
\end{align}
We obtain the functions for the spectrum and the transverse polarization of emitted $e^\pm$,
\begin{align}
F_\mathrm{IS}(x) &= \frac{1}{6}\left(-2x^2+3x-x_0^2\right)-\frac{2\beta}{A}\left(1-x\right)x_0, \\
F_\mathrm{AS}(x)&= \frac{\xi}{6}\sqrt{x^2-x_0^2}\left(2x-2+\sqrt{1-x_0^2}\right), \\
F_\mathrm{T_1}(x) &=-\frac{1}{6}\left(1-x\right)x_0+\frac{2\beta}{3A}\left(x-x_0^2\right), 
\label{eq:FT1}\\
F_\mathrm{T_2}(x)&=\frac{2\beta'}{3A}\sqrt{\left(1-x_0^2\right)\left(x^2-x_0^2\right)}.
\end{align}

As can be seen in Eq.(\ref{eq:beta}),
the $g_{RR}^S$ operator can directly interfere with the $g_{LL}^V$ operator in the SM,
and thus it can provide the primary contribution from the new physics beyond the SM.
The other contributions are all quadratic (including all the other operators).
Since $x_0$  in $F_{\rm IS}$ is small $(x_0 = 9.67 \times 10^{-3})$ due to $m_e \ll m_\mu$,
it is important to observe $P_{\rm T_1}$ and $P_{\rm T_2}$ to extract the primary effect
in the muon decay.
The analysis of the current experimental results at PSI shows \cite{Danneberg:2005xv,Fetscher:2021ldh,Gagliardi:2005fg}
\begin{align}
\frac{\beta}{A} &= \left(1.1 \pm 3.5 \ (\rm{statistical}) \pm 0.5\  ({\rm systematic})\right) \times 10^{-3} , \\
\frac{\beta'}{A} &= \left(-1.3 \pm 3.5 \ (\rm{statistical}) \pm 0.6\  ({\rm systematic})\right) \times 10^{-3} ,
\end{align}
if $g_{RR}^S$ is the only source of the new physics.
The experimental accuracy has not yet reached the size of the SM contribution in $P_{\rm T_1}$.
Though the current experimental bounds for $\beta$, $\beta'$ are loose yet,
we expect that the hundred-times-intense
new muon beamlines in the world can develop the observation of the transverse positron polarization
for ${\beta/A,\beta'/A} \sim 10^{-3}$.
The sensitivity of the order of $10^{-3}$ of those values in the near-future experiment at J-PARC
can be also expected from the simulation in
Ref.\cite{Fukuyama:2019jiq}.

\section{Mu-to-$\mubar$ transition}
\label{sec3}

The effective Mu-to-$\mubar$ transition operators in the Lagrangian is given as \cite{Conlin:2020veq}
\begin{align}
-{\cal L}_{{\rm Mu}-\mubar}  & = \frac{4G_F}{\sqrt2} \left[
  g_1 (\bar \mu \gamma^\alpha P_L e)  (\bar \mu \gamma_\alpha P_L e) + 
g_2 (\bar \mu \gamma^\alpha P_R e)  (\bar \mu \gamma_\alpha P_R e) \right. \\
& \left. \ \ \ \ \ \ \  + g_3(\bar \mu \gamma^\alpha P_L e)  (\bar \mu \gamma_\alpha P_R e) +
g_4 (\bar \mu P_L e)  (\bar \mu P_L e) + 
 g_5 (\bar \mu P_R e)  (\bar \mu P_R e) \right] + {\rm H.c}. \nonumber
\end{align}
There are four states $(F, m) = (0,0)$, $(1,0)$, and $(1,\pm 1)$ in the $1S$ orbital of Mu.
The transition amplitudes for the $(F,m)$ states are 
\begin{align}
{\cal M}_{0,0} &=- \frac{8 (m_{\rm red}\alpha_{\rm em})^3 }{\sqrt2 \pi} \left( g_1 + g_2 - \frac32 g_3 - \frac 14 g_4 - \frac14 g_5\right) G_F,
\label{eq:amplitude00} \\
{\cal M}_{1,m} &= 
- \frac{8 (m_{\rm red}\alpha_{\rm em})^3}{\sqrt2 \pi} 
\left( g_1 + g_2 + \frac12 g_3 - \frac 14 g_4 - \frac14 g_5\right) G_F,
\label{eq:amplitude10}
\end{align}
where $m_{\rm red} = m_e m_\mu/(m_e+m_\mu) \simeq m_e $ is the reduced mass between the muon and electron,
and $\alpha_{\rm em}$ is the fine structure constant.

The external magnetic field mixes the $(0,0)$ and $(1,0)$ states, and the amplitudes in the magnetic flux density $B$
are given as \cite{Horikawa:1995ae,Hou:1995np}
\begin{align}
{\cal M}_{0,0}^B &= \frac12 \left( {\cal M}_{0,0} - {\cal M}_{1,0} + \frac{{\cal M}_{0,0} + {\cal M}_{1,0}}{\sqrt{1+X^2}} \right), \\
{\cal M}_{1,0}^B &= \frac12 \left( -{\cal M}_{0,0} + {\cal M}_{1,0} + \frac{{\cal M}_{0,0} + {\cal M}_{1,0}}{\sqrt{1+X^2}} \right),
\end{align}
where $X = 6.31 \times B/{\rm Tesla}$.
On the other hand, the magnetic field splits the $(1,\pm 1)$ states,
and the oscillations of the $(1,\pm 1)$ states are dropped in the magnetic field
for $B \agt 0.01$ Tesla.

The time-integrated transition probability at the PSI experiment is given as
\begin{equation}
P = 2 \tau^2 \left( |c_{0,0}|^2 | {\cal M}_{0,0}^B |^2 +  |c_{1,0}|^2 | {\cal M}_{1,0}^B |^2
\right),
\label{transition-probability}
\end{equation}
where $|c_{F,m}|^2$ gives the population of the Mu states,
and $\tau$ is the Mu lifetime.
The result by the PSI experiment at the magnetic flux density $B = 0.1$ Tesla is \cite{Willmann:1998gd}
\begin{equation}
P < 8.3 \times 10^{-11}.
\label{P-bound}
\end{equation}
If $g_3 =0$, we obtain
\begin{align}
P &= \frac{64 m_{\rm red}^6 \alpha^6 \tau^2 G_F^2}{\pi^2} \left|{g_1 + g_2 - \frac14 g_4 -\frac14 g_5}\right|^2 \frac{|c_{0,0}|^2+|c_{1,0}|^2}{1+X^2}. 
\end{align}
The PSI experimental result 
 is decoded as
\begin{equation}
\left|g_1 + g_2 - \frac14 g_4 - \frac14 g_5\right| < 3.0 \times 10^{-3}   .
\label{PSI-bound}
\end{equation}
If $g_3 \neq 0$ and the others are zero,
we find 
\begin{equation}
|g_3 | < 2.1 \times 10^{-3} .
\label{G3_bound}
\end{equation}
We use the population of Mu states, $|c_{0,0}|^2 = 0.32$, $|c_{1,0}|^2  = 0.18$.

\bigskip

The Mu-to-$\overline{\rm Mu}$ transition operators 
 are generated at the tree level by the following mediators~\cite{Fukuyama:2021iyw}:
\begin{enumerate}

\item Neutral flavor gauge boson ($\rightarrow g_1, g_2, g_3$)

\item Neutral scalar in an inert $SU(2)_L$ doublet ($\rightarrow g_3, g_4, g_5$)

\item Doubly-charged scalar in the $SU(2)_{L,R}$ triplet ($\rightarrow g_1, g_2$)

\item Dilepton gauge boson in a $SU(3)_l \times U(1)_X$ extension of electroweak symmetry ($\rightarrow g_3$)

\end{enumerate}
If the Mu-to-$\mubar$ transition is generated,
 new muon decay operators are also induced,
especially for the $g_1$ and $g_3$ terms.
For the $g_2$, $g_4$, and $g_5$ terms, 
new muon decays are induced
if the right-handed neutrinos are lighter than muon.
Through the mixings of the left- and right-handed neutrinos,
muons can also decay to active neutrinos plus an electron.

We note that the four-fermion operators induced by the mediators can modify the decay constant,
$\bar G_F$ in Eq.(\ref{eq:decay_rate}),
and therefore, 
the universality of the decay constants can constrain the masses of the mediators and 
their flavor-dependent couplings.
However, 
the Mu-to-$\mubar$ transition (if induced) can provide stronger experimental constraints on the couplings
 than the electroweak precision and high energy Bhabha scattering.
Corrections to the muon decay operators to induce the transverse positron polarization in the $\mu^+$ decay
for our target 
are about $10^{-3}$, and they can be a deeper probe than the current experiments.
Therefore, we do not describe the current experimental constraints from the electroweak precisions in this paper.
See Refs.\cite{Cuypers:1996ia,Falkowski:2015krw,Li:2019xvv,Crivellin:2020klg,Crivellin:2021njn} for the experimental constraints.
We also note that
the contributions to muon and electron $g-2$ are too small to explain their anomalies 
as a consequence of the $10^{-3}$ size of the induced coupling in our context.

In the next section, 
we will consider the model of the neutral flavor gauge boson,
and we show that the Mu-to-$\mubar$ transition and the correction
to $P_{\rm T_1}$ in the polarized $\mu^+$ decay are related.
In the subsequent section, we will also study the other models.

\section{Neutral flavor gauge boson}
\label{sec4}

The interactions to the neutral gauge boson $X$ to generate the Mu-to-$\mubar$ transition are written as \begin{equation}
{\cal L} = g_X (\overline{\ell_\mu} \gamma_\alpha \ell_e + \overline{\ell_e} \gamma_\alpha \ell_\mu) X^\alpha
+ a g_X  (e^{-i \varphi} \overline{ \mu_R} \gamma_\alpha e_R + e^{i \varphi} \overline{ e_R} \gamma_\alpha \mu_R) X^\alpha,
\label{NGB-lagrangian}
\end{equation}
where
$\ell_e$ and $\ell_\mu$ are the left-handed lepton doublets:
\begin{equation}
\ell_e = \left( 
 \begin{array}{c}
 \nu_{eL} \\
 e_L 
 \end{array}
\right), 
\qquad
\ell_\mu = \left( 
 \begin{array}{c}
 \nu_{\mu L} \\
 \mu_L 
 \end{array}
\right), 
\end{equation}
and
$a$ is a $U(1)'$ charge for the right-handed charged leptons.
In Appendix \ref{appendix:ref}, 
we give a construction of this model.
Here, we only mention that the interactions have a discrete lepton flavor symmetry,
and they do not induce the $\Delta L_e, \Delta L_\mu = \pm 1$ processes,
such as $\mu \to e\gamma$, $\mu \to 3e$. 
See Ref.\cite{Fukuyama:2021iyw} for the bound of the $\Delta L_e, \Delta L_\mu = \pm 1$ processes 
in the case where the discrete symmetry is not exact. 
There can be a physical phase parameter $\varphi$ in the coupling in general.
The phase in the left-handed lepton couplings can be rotated away without loss of generality.

The Mu-to-$\mubar$ transition operators can be generated by the exchange of the neutral gauge boson, and
we obtain
\begin{equation}
g_1 = \frac{g_X^2}{4 \sqrt2 M_X^2 G_F},
\qquad
g_2 = a^2  e^{-2i \varphi} \frac{g_X^2}{4 \sqrt2 M_X^2 G_F},
\qquad
g_3 = 2a e^{-i \varphi} \frac{g_X^2}{4 \sqrt2 M_X^2 G_F},
\end{equation}
where $M_X$ is the mass of the neutral gauge boson.

The interaction can also generate the following eEDM
via a loop diagram with the muon mass insertion in the internal line:
\begin{equation}
\frac{d_e}{e} = m_\mu \frac{2a g_X^2 \sin \varphi}{64 \pi^2 M_X^2} G\left(\frac{m_\mu^2}{M_X^2} \right)
= \frac{m_\mu G_F {\rm Im}\, g_3^* }{8\sqrt2\pi^2} G\left(\frac{m_\mu^2}{M_X^2} \right),
\end{equation}
where $G$ is a loop function,
\begin{equation}
G (x) = \frac{4-3x -x^3 + 6x \ln x}{(1-x)^3}.
\end{equation}
The experimental bound of the eEDM is \cite{ACME:2018yjb}
\begin{equation}
|d_e| < 1.1 \times 10^{-29} \ e\cdot {\rm cm}.
\end{equation}
One needs $|\varphi| \alt 10^{-5}$
if we consider the $|g_3| \sim 2\times 10^{-3}$ region which is near the current bound from the Mu-to-$\mubar$ 
transition experiment in Eq.(\ref{G3_bound}).
Therefore, we suppose that there is no CP phase in the interaction, $\varphi=0$.

The exchange of the neutral gauge boson can also induce the muon decay operators.
Using the Fierz transformation,
\begin{equation}
(\overline{ \nu_{\mu L}} \gamma  \nu_{e L}) (\overline{e_R} \gamma  \mu_R) 
= - 2 (\overline{e_R} \nu_e) ( \overline{ \nu_\mu } \mu_R),
\end{equation}
we find
\begin{equation}
g_{RR}^S = -2 {g_3}.
\end{equation}
Similarly, the exchange can induce the correction of $g_{LL}^V$,
\begin{equation}
\Delta g_{LL}^V = g_1.
\end{equation}
If the neutrinos are Majorana fermions, we also find
\begin{equation}
g_{RR}^V = -g_3, \qquad g_{LL}^S = 2 g_1.
\end{equation}
From the definitions of the muon decay parameters $\beta$ in Eq.(\ref{eq:beta}),
we obtain
\begin{equation}
\beta = 4 ( + 2 g_1 g_3 + 2 g_3 (1+ g_1))   \simeq 8 g_3.
\end{equation}

\begin{figure}[t]
\center
\includegraphics[width=8.8cm]{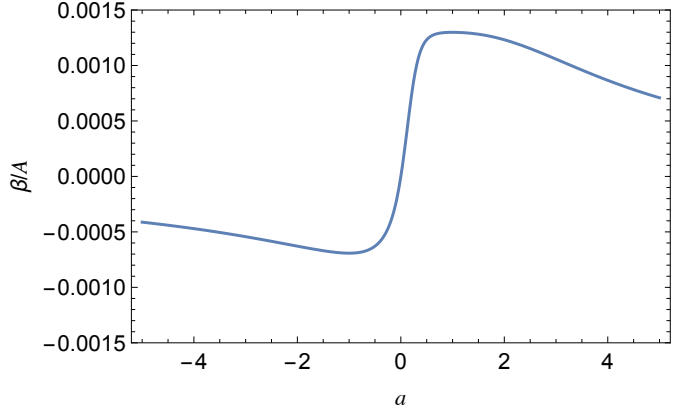}
\caption{
We plot the muon decay parameter $\beta/A$ as a function of the model parameter $a$,
supposing that the probability of the Mu-to-$\mubar$ transition is just at the current experimental bound.
The transition experiment allows a larger magnitude of $\beta/A$ for positive values than for negative values.
}
\label{fig1}
\end{figure}

\begin{figure}[t]
\center
\includegraphics[width=8.8cm]{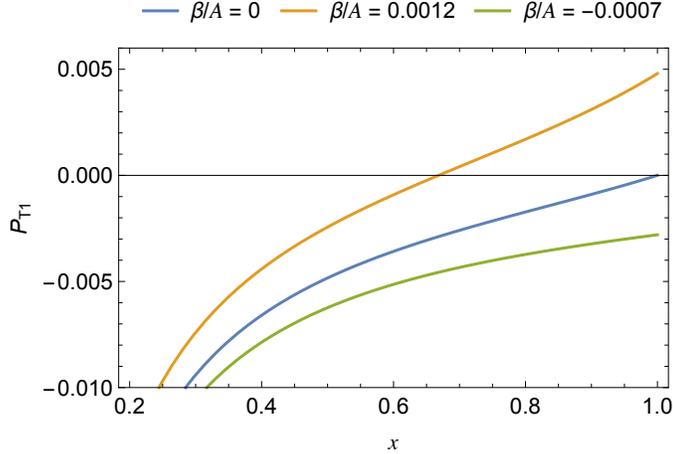}
\caption{
We plot $P_{\rm T_1}(\theta = \pi/2)$ as a function of the reduced positron energy, $x = E_e/W_{e\mu}$,
for various $\beta/A$.
The SM case corresponds to $\beta/A =0$.
}
\label{fig2}
\end{figure}

The model parameters are constrained by 
the time-integrated transition probability (with $B = 0.1$ T) given in Eq.(\ref{transition-probability}).
Since there are no $\Delta L_e$, $\Delta L_\mu = \pm 1$ processes in the model in Eq.(\ref{NGB-lagrangian})
and the Mu-to-$\mubar$ transition gives the strongest constraints,
the experimental constraints for the model parameters $g_X$, $M_X$ and $a$ are 
governed by the bound of the transition probability in Eq.(\ref{P-bound}).
We assume that the neutral gauge boson is heavier than the muon.
In future, Belle II (the ILC) can directly search for the boson with $M_X\lesssim 10$ GeV ($\mathcal{O}(100)$ GeV) if its integrated luminosity accumulates enough.
The transition probability
is proportional to
\begin{equation}
|c_{0,0}|^2 \left| - 1.68 g_3 + {g_1+ g_2 } \right|^2
+
|c_{1,0}|^2 \left| 0.68 g_3+ {g_1+ g_2}\right|^2.
\end{equation}
Notice that $g_1$ and $g_2$ are positive, and $g_3$ can be either positive and negative.
One can find that
the magnitude of $g_3$ allowed by the transition experiment 
depends on the sign of $g_3$.
Indeed, the larger magnitude is allowed for $g_3 >0$.
In Fig.\ref{fig1}, we show a plot of $\beta/A$
as a function of $a$,
assuming that
the transition probability is just the upper bound from the PSI experiment.

In Fig.\ref{fig2},
we plot $P_{\rm T_1}$ given in Eq.(\ref{eq:PT1}).
We choose $\beta/A = 0.0012$ and $\beta/A = -0.0007$ 
which are allowed by the experimental result of the Mu-to-$\mubar$ transition.
The electron mass can induce $P_{\rm T_1}$ in the SM case, $\beta/A = 0$.
Around the maximal energy of the positron, $P_{\rm T_1}$ tends to zero.
As can be found from Eq.(\ref{eq:FT1}),
$P_{\rm T_1}$ changes its sign for $\beta > 0$ for larger energy.
We note that the differential decay width given in Eq.(\ref{eq:decay_rate})
is larger for the larger positron energy,
and the near-future experiment may observe the change of the sign.

\section{Other models}
\label{sec5}

As examined in the previous section,
the flavor neutral gauge boson can generate both
the Mu-to-$\mubar$ transition and the transverse positron polarization $P_{\rm T_1}$ in the muon decay,
and those two are related.
In this section, we study other models
to generate the new muon decay operators
and see if the transverse polarizations are related with the Mu-to-$\mubar$ transition.

We first enumerate the interactions to generate the following new muon decay operators at the tree level:
\begin{align}
{\rm (a)} & \quad
h_{ee} \overline{e_R} \ell_e \Phi + h_{\mu\mu} \overline{\mu_R} \ell_\mu \Phi + {\rm H.c.},
\label{eq.a}
\\
{\rm (b)} & \quad
h_{e\mu} \overline{e_R} \ell_\mu \Phi + h_{\mu e}\overline{\mu_R} \ell_e \Phi + {\rm H.c.},
\\
{\rm (c)} & \quad
\kappa_{ee} \overline{\ell_e^c} \ell_e \Delta_L + \kappa_{\mu\mu} \overline{\ell_\mu^c} \ell_\mu \Delta_L + {\rm H.c.},
\\
{\rm (d)} & \quad
\kappa_{e\mu} (\overline{\ell_e^c} \ell_\mu \Delta_L +\overline{\ell_\mu^c} \ell_e \Delta_L) + {\rm H.c.},
\\
{\rm (e)} & \quad
f(\overline{\ell_e^c} \ell_\mu S^+ - \overline{\ell_\mu^c} \ell_e S^+) + {\rm H.c.},
\\
{\rm (f) }& \quad
g_{3l} (\overline{(e_R)^c} \gamma_\alpha \ell_e Y^\alpha + \overline{(\mu_R)^c} \gamma_\alpha \ell_\mu Y^\alpha) + {\rm H.c.},
\\
{\rm (g)} & \quad
g_{3l} (\overline{(e_R)^c} \gamma_\alpha \ell_\mu Y^\alpha + \overline{(\mu_R)^c} \gamma_\alpha \ell_e Y^\alpha) + {\rm H.c.},
 \\
{\rm (h)} & \quad
g_X (\overline{\ell_e} \gamma_\alpha \ell_\mu X^\alpha +  a \,\overline{e_R} \gamma_\alpha \mu_R X^\alpha) + {\rm H.c.}
\label{eq.h}
\end{align}
Here, $\Phi$ is a $SU(2)_L$ inert doublet which does not have a 
vacuum expectation value (vev),
$\Delta_L$ is a $SU(2)_L$ triplet whose vev can generate the type-II neutrino masses \cite{Schechter:1980gr,Cheng:1980qt,Lazarides:1980nt,Mohapatra:1980yp},
and $S^+$ is a $SU(2)_L$ singlet with hypercharge $Y=1$.
The couplings to $\Delta_L$ and $S^+$ are written in terms of the components as
\begin{align}
\overline{\ell_a^c} \ell_b \Delta_L &= \overline{(\nu_{aL})^c} \nu_{bL} \Delta_L 
- \frac1{\sqrt2} \left( \overline{(\nu_{aL})^c} e_{bL} + \overline{(e_{aL})^c } \nu_{bL} \right) \Delta_L^+
+ \overline{(e_{aL})^c} e_{bL} \Delta^{++}_L, \\
\overline{\ell_a^c} \ell_b S^+ &=
  \left( \overline{(\nu_{aL})^c} e_{bL} - \overline{(e_{aL})^c } \nu_{bL} \right) S^+.
\end{align}
The vector field $Y_\alpha = (Y_\alpha^{++},Y_\alpha^+)$ denotes a multiplet of the dilepton gauge boson in a model with
gauge extension, and $X_\alpha$ is a flavor neutral gauge boson, which we have studied in the previous section.
The coexistence of (a) and (b) suffers from the LFV decay constraints if the couplings are sizable.
The same is true for the coexistence of (c) and (d), and that of (f) and (g).
A discrete flavor symmetry can forbid the coexistence of them.
For example, we assign the discrete charges $c_i$ to the lepton fields and the SM Higgs $H$ as
\begin{equation}
\ell_e, e_R : c_1, \qquad \ell_\mu, \mu_R : c_2, \qquad  \ell_\tau, \tau_R : c_3, \qquad H : 0,
\label{discrete-sym}
\end{equation}
and $c_i$'s are all different.
If the charge of $\Phi$ is 0, case (a) is obtained.
If we assign the discrete charge $n$ to $\Phi$ under $Z_{2n}$ symmetry
and $c_1 - c_2 \equiv n$,
we obtain case (b).
In fact, $\Phi$ and $H$ should not mix in case (b) to suppress the LFV,
which can be also controlled by the discrete symmetry.
For (e), it also suffers from the LFV if $S^+$ also couples to $\tau$.
These are all the lepton bilinear couplings (without right-handed neutrinos)
that cause muon decays at the tree level.
The muon decay operators from the couplings with the right-handed neutrinos 
are suppressed by the heavy-light neutrino mixings, which will be studied in the context of the
left-right model later.

In Table \ref{Table1}, we list the muon decay operators which can be induced by the interactions 
from (a) to (h).
If the induced muon decay process is
$\mu \to e \nu_e \overline{\nu_\mu}$, which does not interfere with 
$\mu \to e \overline{\nu_e} \nu_\mu$,
the Fierz-transformed operator given in Appendix \ref{appendix:A} is shown by assuming that the neutrinos are Majorana.
One can see that
$g_{RR}^S$ can be induced 
in the cases (a), (g), and (h),
and $P_{\rm T_1}$ can be modified from the SM since
$\beta/A$ can be $\sim 10^{-3}$.
Due to the constraint of the eEDM,
only case (a) can generate $\beta'/A \sim 10^{-3}$,
and $P_{\rm T_2}$ can be observed in the near-future experiments.
Only case (h) can relate the transverse polarization
and the Mu-to-$\mubar$ transition as discussed in the previous section.

\begin{table}
\caption{
We list which muon decay operators are induced from
the respective interactions and mediators given in Eqs.(\ref{eq.a})-(\ref{eq.h}).
We assume that the neutrinos are Majorana
to make it interfere with the SM decay operator  
if the induced operator is for
$\mu \to \bar\nu_\mu \nu_e e$ (See Appendix \ref{appendix:A}).
In the fourth column, we put ``Im"
if the phase of the coefficient is allowed.
The $\sharp$ mark is attached
if the phase is constrained from the existence of the $(\Phi \tilde H)^2$ term
in the scalar potential (See explanations in the text).
If the eEDM constrains the phase, we put ``Re ($\because$ eEDM)".
In the fifth column, we put the transition operators if they are induced.
}
\center
\begin{tabular}{c|c|c|c|c}
interaction & mediator & operator & phase & Mu-to-$\overline{\rm Mu}$\\
\hline
(a) & $\Phi^+$ & $g_{RR}^S$ & Im & \\
(b) & $\Phi^+$ & $g_{RR}^V$ & Im ($\sharp$) & $g_3$ \\
(c) & $\Delta_L^+$ & $g_{LL}^S$  & Im & $g_1$ \\
(d) or (e) & $\Delta_L^+$ or $S^+$ & $g_{LL}^V$   & Im & \\
(a)+(c) & $\Phi^+$--$\,\Delta_L^+ $ & $g_{LR,RL}^S$   & Im & $g_1$ \\
(b)+((d) or (e)) & $\Phi^+$--$\,(\Delta_L^+$ or $S^+) $& $g_{LR,RL}^{S,T}$  &Im & $g_3$ \\
(b)+(d) & $\Phi^0$--$\,\Delta_L^0$ & $g_{LR,RL}^{S,T}$  &Im  & $g_3$ \\
(b)+(d) & $\Phi^{0*}$--$\,\Delta_L^0$ & $g_{LR,RL}^{V}$  &Im & $g_3$ \\
(f) & $Y^+$ & $g_{RR}^{V}$  & Im & $g_3$ \\
(g) & $Y^+$ & $g_{RR}^{S}$ & Re ($\because$ eEDM) &  \\
(h) & $X$ & $g_{RR}^{S,V}, g_{LL}^{S,V}$ & Re ($\because$ eEDM) & $g_1$,$g_2$,$g_3$ \\
\end{tabular}
\label{Table1}
\end{table}

\subsection{Inert Higgs doublet $\Phi$}

The interaction (a) can directly induce $g_{RR}^S$
via $\Phi^+$ exchange,
\begin{equation}
g_{RR}^S \propto h_{ee} h_{\mu\mu}^*.
\end{equation}
The EDMs for electron and muon are obtained by a $\Phi^0$ loop diagram,
and
\begin{equation}
d_a \propto m_a\, {\rm Im} h_{aa}^2 
\left( f(m_a, M_{\rm Re}^2) - f(m_a,M_{\rm Im}^2)  \right) \quad (a = e, \mu),
\end{equation}
where $M_{\rm Re}$ and $M_{\rm Im}$ are the masses of the real and imaginary parts of $\Phi^0$,
and $f$ is a loop function.
One can find that $g_{RR}^S$ can be complex without contradicting EDMs for electron and muon
if $h_{ee}^2$ is real or $M_{\rm Re} = M_{\rm Im}$.
The magnitude of the muon EDM ($\mu$EDM) for $P_{\rm T_2} \sim O(10^{-3})$ in the case of
imaginary $h_{\mu\mu}$ and $M_{\rm Re} \neq M_{\rm Im}$ is estimated to be 
$O(10^{-24})\ e \cdot {\rm cm}$, which is far below the current experimental bound \cite{Muong-2:2008ebm,Ema:2021jds}.
The absence of $(\Phi \tilde H) (\Phi \tilde H)$ term ($H$ is a Higgs doublet which acquires a vev) 
can make $M_{\rm Re} = M_{\rm Im}$,
though discrete symmetries cannot realize it in non-supersymmetric models.
We note that the size of the $(\Phi \tilde H) (\Phi \tilde H)$ term
is related with the radiative neutrino mass with the inert doublet \cite{Ma:2006km}.
Anyway, $g_{RR}^S$ can be complex, and therefore,
both $P_{\rm T_1}$ and $P_{\rm T_2}$ can be observed in the near-future experiment in this case.
The Mu-to-$\mubar$ transition is not induced.

In case (b),
the operator which is induced by the charge scalar $\Phi^+$ exchange
is
\begin{equation}
(\overline{e_R} \nu_\mu  )( \overline{\nu_e}  \mu_R),
\label{eq}
\end{equation}
which is not $g_{RR}^S$.
If the neutrinos are Majorana, 
the induced operator can become $g_{RR}^V$ (instead of $g_{RR}^S$).
We find that the relation of the Mu-to-$\mubar$ transition is
\begin{equation}
g_{RR}^V \simeq - g_3^* \propto h_{e\mu} h_{\mu e}^*.
\end{equation}
The eEDM is
\begin{equation}
d_e \propto m_\mu \, {\rm Im} (h_{\mu e} h_{e\mu})
\left( f(m_\mu, M_{\rm Re}^2) - f(m_\mu, M_{\rm Im}^2)  \right).
\end{equation}
The coupling of $(\Phi \tilde H) (\Phi \tilde H)$ term and
only one of $h_{\mu e}$ and $h_{e\mu}$ can be made to be real 
by the redefinition of the lepton fields and $\Phi$.
 Therefore,
in order to make Im$\, g_{RR}^V \neq 0$
in agreement with the eEDM,
$M_{\rm Re} = M_{\rm Im}$ is needed in this case.
We note that the eEDM diagram hits the muon mass at the internal line,
while the electron mass is hit for the $\mu$EDM and thus the $\mu$EDM becomes much smaller than the eEDM.

If we take into account the light-heavy neutrino mixings,
$g_{RR}^S$ can be induced even in the case (b).
The current neutrino state can be written by the mass eigenstates as
\begin{equation}
\nu_a  = U_{a i} \nu_i + X_{a I} N_I.
\end{equation}
See Appendix \ref{appendix:B} for the neutrino mixing matrix.
We find
\begin{equation}
\{ \beta, \beta^\prime \} \sim \{ {\rm Re} {g_3}, {\rm Im} g_3 \} \sum_{i,j} U_{e i}  U^*_{\mu i} U^*_{e j} U_{\mu j} .
\end{equation}
We have supposed that $N_I$'s are heavier than muon.
Using the unitarity relation, we find
\begin{equation}
\sum_i U_{e i} U^*_{\mu i} = - \sum_{I} X_{e I} X_{\mu I}^*. 
\end{equation}
This magnitude is constrained by the $\mu \to e \gamma$ decay process,
and the transverse positron polarizations are tiny in the case (b).

\subsection{Type-II seesaw}

The interactions (c) and (d) are available for type-II seesaw neutrino masses
when the $SU(2)_L$ triplet $\Delta_L$ acquires a vev.

In case (c),
the $g_{LL}^S$ muon decay operator is generated
by $\Delta_L^+$ exchange,
if the neutrinos are Majorana.
The Mu-to-$\mubar$ transition operator ($g_1$) is also generated by
$\Delta_L^{++}$ exchange:
\begin{equation}
g_{LL}^S = 2 {g_1^*}.
\end{equation}
In case (d), the $g_{LL}^V$ contribution of the muon decay is generated,
while it does not induce the Mu-to-$\mubar$ transition.
The type-II seesaw interactions do not generate the EDMs,
and thus, the induced coefficients can be imaginary.

\subsection{Type-II seesaw + inert doublet $\Phi$} 

The $\beta$, $\beta'$ parameters for the transverse polarizations
in Eqs.(\ref{eq:beta}) and (\ref{eq:beta-prime})
are not generated from the type-II seesaw terms alone.
If we add the inert doublet $\Phi$
and there are multiple contributions (b)+(c),
$\beta$, $\beta'$ can be generated
and they relate with the Mu-to-$\mubar$ transitions as follows:
\begin{equation}
\{ \beta, \beta' \} = \{ {\rm Re}, {\rm Im} \} (8 g_1^* g_3).
\end{equation}
Those magnitudes are less than $O(10^{-5})$ from the PSI bound of the Mu-to-$\mubar$ transition.

The scalar trilinear term $\Phi H \Delta_L$ is allowed, and it can induce a $\Phi$-$\Delta_L$ mixing
since the SM Higgs doublet $H$ acquire a vev.
Then, $g_{LR,RL}^{S,V,T}$ operators can be generated (See Table 1).
Among them,
$g_{LR,RL}^V$ can be generated by 
the neutral scalar exchange with $(\Phi \tilde H) (\Phi \tilde H)$ insertion 
in the interaction with (b)+(d).
The generated operators are
\begin{equation}
g_{LR}^{S,T} \propto \kappa_{e\mu}^* h_{\mu e}^*, \quad
g_{RL}^{S,T} \propto \kappa_{e\mu} h_{e\mu}, \quad
g_{LR}^{V} \propto \kappa_{e\mu} h_{\mu e}^*, \quad
g_{RL}^{V} \propto \kappa_{e\mu}^* h_{e\mu}.
\end{equation}
We note that the coupling of $\Phi H \Delta_L$ can be made to be real
by the phase redefinition of $\Delta_L$.
We obtain the CP violating parameter $\alpha'$ in Eq.(\ref{eq:alpha-prime}) for $P_{\rm T_2}$ as
\begin{equation}
\alpha'  \propto  |\kappa_{e\mu}|^2 \, {\rm Im}  (h_{\mu e} h_{e\mu} ).
\end{equation}
As explained, the existence of the $(\Phi \tilde H) (\Phi \tilde H)$ term
can conflict with the eEDM, and $h_{\mu e}$ and $h_{e\mu}$ should be real.
As a consequence, 
the muon decay parameter $\alpha'$ is severely constrained by the eEDM
in this model.

\subsection{Dilepton gauge boson}

In the dilepton gauge model \cite{Frampton:1989fu,Frampton:1992wt,Frampton:1991mt,Frampton:1997in,Fujii:1992np,Fujii:1993su}
whose gauge symmetry is $SU(3)_c \times SU(3)_l \times U(1)_X$,
the leptons are unified in one multiplet, ${\bf 3}^*$ representation of $SU(3)_{l}$,
$L_a = (l_a, - \nu_a, l^c_a)$.
The gauge interaction (in two-component spinor notation) of the dilepton gauge boson is given as
\begin{equation}
{\cal L} = g_{3l} ( \nu_a \sigma^\mu \overline{l^c_a}Y^+_\mu - l_a \sigma^\mu  \overline{l^c_a} Y^{++}_\mu ) + {\rm H.c.}
\end{equation}
The Higgs boson to generate the charged lepton masses are ${\bf 3}^*$ and ${\bf 6}$ under $SU(3)_l$.
Remind that the coupling matrices with ${\bf 3}^*$ and $\bf 6$ are anti-symmetric and symmetric, respectively, under the generation index $a$.
If the Yukawa couplings with the ${\bf 3}^*$ Higgs boson are absent,
the mass matrix is symmetric and the gauge interaction of the mass eigenstates is given by (f).
By adopting a discrete flavor symmetry, e.g.,
$L_1 : 1$, $L_2 : 2$, $L_3 : 0$, ${\bf 3}^*$ and ${\bf 6}$ : 0
under $Z_3$,
the allowed Yukawa couplings can be
\begin{equation}
L_1 L_2 \, {\bf 3}^* - L_2 L_1 \, {\bf 3}^* + L_1 L_2 {\bf 6} + L_2 L_1 {\bf 6} + L_3 L_3 {\bf 6}.
\end{equation}
In this case, the gauge interaction is given by (g),
because the multiplets are $L_a = (e,-\nu_e,\mu^c)$, $(\mu,-\nu_\mu,e^c)$, and $(\tau,-\nu_\tau,\tau^c)$.
In order to make $m_e \ll m_\mu$, one needs a fine-tuning.

In case (f),
the Mu-to-$\mubar$ transition operator $g_3$ is generated by the $Y^{++}$ exchange.
The $g_{RR}^V$ muon decay operator can be generated by the $Y^+$ exchange,
\begin{equation}
g_{RR}^V = - g_3^*,
\end{equation}
assuming that the neutrinos are Majorana.

In case (g),
the $g_{RR}^S$ muon decay operator is generated by the $Y^+$ exchange.
Therefore, the modification of $P_{\rm T_1}$ from the SM can be sizable to detect in the muon decay experiments.  
In general, the couplings with $Y^+$ are complex in the basis where the charged lepton masses are real.
However, the complex couplings can induce the eEDM via the $Y^{++}$ loop diagram.
Due to the eEDM bound, the phase of $g_{RR}^S$ has to be tiny,
and $P_{\rm T_2}$ will not be observed in the near-future experiment.

\subsection{Left-right model}
\label{sec:left-right}

The involvement of the right-handed neutrinos 
also provides muon decay operators which can interfere with the SM decay amplitude.
Though their contributions are small due to the heavy-light neutrino mixings as we mentioned,
we describe the contributions to $g_{RR}^S$ in the left-right model,
$SU(2)_L \times SU(2)_R \times U(1)_{B-L}$ gauge theory, 
as a pedagogical guide.

We introduce a $SU(2)_R$ triplet $\Delta_R$
and consider the interaction,
\begin{equation}
\overline{\ell_{eR}^c} \ell_{eR} \Delta_R + \overline{\ell_{\mu R}^c} \ell_{\mu R} \Delta_R + {\rm H.c.},
\end{equation}
where $\ell_R$ is a $SU(2)_R$ doublet, e.g.,  $\ell_{eR} = (N_{eR}, e_R)^T$.
The vev of the $SU(2)_R$ triplet breaks $SU(2)_R \times U(1)_{B-L}$ down to $U(1)_Y$,
and also generates the Majorana masses of the right-handed neutrinos.
The $\Delta_R^{++}$ exchange generates the $g_2$ operator of the Mu-to-$\mubar$ transition.
The $\Delta_R^+$ exchange generates
\begin{equation}
(\overline{e_R} (N^c_e)_L ) (\overline{ (N^c_\mu)_L} \mu_R).
\end{equation}
We use the notation of the neutrino mixing matrix given in Appendix \ref{appendix:B},
and the current neutrino state can be written by the mass eigenstates as
\begin{equation}
(N^c_\alpha)_L   = V_{\alpha i} \nu_i + Y_{\alpha I} N_I.
\end{equation}
We obtain
\begin{equation}
\{ \beta, \beta' \} = \{ {\rm Re}, {\rm Im} \} \left(-8 {g_2^*} \sum_{i,j} U_{ei} V_{ei}^* U^*_{\mu j} V_{\mu j} \right).
\end{equation}
We remark that the magnitude of $U_{ei} V_{ei}^*$ is directly constrained by the 
neutrinoless double beta decay ($0\nu2\beta$).
The unitarity and the decay universality restrict $U_{\mu i} V_{\mu i}^* = - X_{\mu I} Y_{\mu I}^*$
(See Appendix \ref{appendix:B}).
The induced size of $|\beta^{{(\prime)}}|$ is estimated to be less than $O(10^{-7})$.

The tree-level $W_R$ gauge boson exchange 
can generate the $g_{RR}^S$ operator if the neutrinos are Majorana \cite{Doi:1981sn}.
Using 
\begin{equation}
(\overline{e_R} \gamma N_{eR}) (\overline{N_{\mu R}} \gamma \mu_R)
= 2(\overline{e_R} (N_{\mu}^c)_L) (\overline{(N_e^c)_L} \mu_R),
\end{equation}
we obtain the muon decay parameter from the $W_R$ exchange as
\begin{equation}
\{ \beta, \beta' \} =  \{ {\rm Re}, {\rm Im} \}
\left( -8 \frac{g_R^2}{g_L^2} \frac{M_{W_L}^2}{M_{W_R}^2} 
\sum_{i,j} U_{ei} V_{\mu i}^* U^*_{\mu j} V_{e j} \right),
\end{equation}
which is also very tiny 
due to the $W_R$ mass bound from the LHC \cite{Nemevsek:2018bbt,Sirunyan:2018pom,CMS:2021dzb,Aaboud:2019wfg}
and $0\nu2\beta$.
For a native estimation of the quantity $|\sum_{i,j} U_{ei} V_{\mu i}^* U_{\mu j}^* V_{ej}|$
 (See Appendix \ref{appendix:B}),
we obtain $|\beta^{{(\prime)}}|$ is less than $O(10^{-8})$.
Even for more conservative estimation of the quantity,
$|\beta^{{(\prime)}}|$ is less than $O(10^{-6})$.

\section{Conclusion}
\label{sec6}

The new leptonic interactions with a discrete flavor symmetry
can induce the Mu-to-$\mubar$ transition and the transverse polarization of $e^{\pm}$
in the polarized $\mu^{\pm}$ decay,
which can be of a size that will be observable at the facilities with high-intensity muon beamlines.
We have studied whether the transition rate and the transverse polarization
can be related.

There are three candidates of the mediators 
to induce the testable muon decay parameter $\beta$ for the transverse positron polarization 
in the near future:
\begin{itemize}
\item
 Neutral flavor gauge boson, 
\item Inert doublet,
\item Dilepton gauge boson.
\end{itemize}
Among them,
in the model of the neutral flavor gauge boson, 
the Mu-to-$\mubar$ transition and the $\beta$ parameter
(the correction of the transverse positron polarization $P_{\rm T_1}$)
are indeed related.
A larger contribution is allowed by the Mu-to-$\mubar$ transition experiment for the 
positive value of $\beta$ than the negative value
($P_{\rm T_1}$ changes its sign depending on the positron energy for positive $\beta$).
The other direction of the transverse polarization, $P_{\rm T_2}$, is constrained by 
the non-observation of the eEDM.

In the model with an inert scalar doublet (which does not acquire a vev),
one of the Mu-to-$\mubar$ transition and the correction to the transverse positron polarization
can be observed.
The non-zero value of $P_{\rm T_2}$ does not conflict with the eEDM in this model.
In the dilepton gauge boson,
one of the the Mu-to-$\mubar$ transition and the correction to the transverse positron polarization
can be observed.
The non-observation of the eEDM restricts $P_{\rm T_2}$.
Though the observable size of the Mu-to-$\mubar$ transition can be induced
in the model with $SU(2)_L$ and $SU(2)_R$ triplet scalars,
the correction to the transverse polarization is smaller than the three above.

\section*{Acknowledgements}

This work was supported in part by the COREnet project of RCNP, Osaka University (T.F.) and JSPS KAKENHI Grant Numbers JP18H01210 and JP21H00081 (Y.U.).

\appendix

\section{Fierz transformations of the muon decay operators for Majorana neutrinos}
\label{appendix:A}

The following identical equations hold for four-component fermions $\psi$ and $\chi$:
\begin{equation}
\overline{\psi} \chi = \overline{\chi^c} \psi^c,
\qquad
\overline{\psi} \gamma_\mu \chi = -\overline{\chi^c} \gamma_\mu \psi^c,
\qquad
\overline{\psi} \sigma_{\mu\nu} \chi = -\overline{\chi^c} \sigma_{\mu\nu} \psi^c,
\end{equation}
where $\psi^c = C \overline{\psi}^T$ and $C$ is a charge conjugation matrix.
Using these, one finds for Majorana neutrinos $\nu = \nu^c$,
\begin{align}
(\overline{e_R} \nu_{aL}) (\overline{\nu_{bL}} \mu_R) &=
-\frac12 (\overline{e_R} \gamma \mu_R) (\overline{\nu_{bL}} \gamma \nu_{aL})
= \frac12 (\overline{e_R} \gamma \mu_R) (\overline{\nu_{aR}} \gamma \nu_{bR}) \nonumber \\
&= \frac12 (\overline{e_R} \gamma \nu_{bR}) (\overline{\nu_{aR}} \gamma \mu_R).
\end{align}
Similarly,
\begin{equation}
(\overline{e_L} \nu_{aR}) (\overline{\nu_{bR}} \mu_L) 
= \frac12  (\overline{e_L} \gamma \nu_{bL}) (\overline{\nu_{aL}} \gamma \mu_L),
\end{equation}
Here, we omit the obvious Lorentz indices of $\gamma_\mu$ for their contraction.
For example, 
though $(\overline{e_R} \nu_{\mu L}) (\overline{\nu_{e L}} \mu_R) $
does not interfere with the usual $\mu \to e \overline{\nu_e} \nu_\mu$ decay amplitude
for Dirac neutrinos,
it can be made to be an operator to interfere with it
for Majorana neutrinos using the above equation.

One can also obtain for $\nu = \nu^c$,
\begin{align}
(\overline{e_L} \gamma \nu_{aL}) (\overline{\nu_{bR}}\gamma \mu_R) 
&=
(\overline{e_L} \gamma \nu_{bL}) (\overline{\nu_{aR}}\gamma \mu_R) ,
\\
(\overline{e_L} \nu_{aR}) (\overline{\nu_{bL}} \mu_R) 
&=
-\frac12(\overline{e_L} \nu_{bR}) (\overline{\nu_{aL}} \mu_R) 
+\frac18(\overline{e_L} \sigma \nu_{bR}) (\overline{\nu_{aL}}\sigma \mu_R), 
\\
(\overline{e_L} \sigma \nu_{aR}) (\overline{\nu_{bL}}\sigma \mu_R) 
&= 
6(\overline{e_L} \nu_{bR}) (\overline{\nu_{aL}} \mu_R) 
+\frac12(\overline{e_L} \sigma \nu_{bR}) (\overline{\nu_{aL}}\sigma \mu_R) ,
\end{align}
and the same for the exchange of $L \leftrightarrow R$.

\section{The model with the neutral flavor gauge boson}
\label{appendix:ref}

We describe the construction of the model with the neutral flavor gauge boson
discussed in Section \ref{sec4}.

Table \ref{Table2} shows extra $U(1)$ charge assignments of the lepton fields.
The extra $U(1)$ symmetries do not cause gauge anomalies:
$[SU(3)_c]^2 U(1)_n$, $[SU(2)_L]^2 U(1)_n$, $[U(1)_Y]^2 U(1)_n$,
$[U(1)_n]^2 U(1)_Y$, $[U(1)_n]^3$, $[U(1)_1]^2 U(1)_2$, $[U(1)_2]^2 U(1)_1$,
and $[{\rm gravity}]^2 U(1)_n$ ($n=1,2$).

\begin{table}[t]
\caption{
We list the $U(1)_1 \times U(1)_2$ charge assignments of the left-handed lepton doublets $\ell_i$,
right-handed charged leptons $e_{iR}$,
and SM singlet scalar fields, $\phi$, $\phi_1$, $\phi_2$.
The scalar field $\phi$ breaks the $U(1)_1 \times U(1)_2$ down to $U(1)'$.
The scalar fields $\phi_1$, $\phi_2$ break the remaining $U(1)'$ symmetry,
and can generate the Yukawa interaction of the first and second generations of the charged leptons.
The $U(1)$ charges of the quark fields are all zero.
}
\center
 \begin{tabular}{|c|c|c|c|c|c|c||c|c|c|}
  \hline 
 fields & $\ell_1$ & $\ell_2$ & $\ell_3$ &
  $e_{1R}$ & $e_{2R}$ & $e_{3R}$ & $\phi$ & $\phi_1$ & $\phi_2$ \\ \hline
  $U(1)_1$ charge & $+1$ & $-1$ & 0 &  $+1$ & $-1$ & 0 & $a_1$ & 1 & 1 \\
  $U(1)_2$ charge & $+1$ & $-1$ & 0 &  $-1$ & $+1$ & 0 & $a_2$ & 1 & $-1$ \\ \hline
  $U(1)'$ charge & $+1$ & $-1$ & 0 &  $a$ & $-a$ & 0 & $0$ & 1 & $a$ 
  \\  \hline
 \end{tabular}
 \label{Table2}
\end{table}

The $\ell_3$ and $e_{3R}$ fields are identified to the third generation,
and the Yukawa interaction to generate the mass of the tau lepton can be directly written.
The Yukawa interaction to generate the electron and muon masses
can be obtained by introducing the vector-like fermions, $L$, $E$,
as in usual flavor models:
\begin{equation}
-{\cal L}_Y = y_1 \phi_1^*  \overline{L_R}  \ell_1 + y_2 \phi_1 \overline{L_R} \ell_2 
+ y'_1 \phi_2^* \overline{e_{1R}} E_L  + y'_2 \phi_2 \overline{e_{2R}} E_L
+ y \overline{E} L H + M_L \overline L L + M_E \overline E E. 
\label{Lag-f}
\end{equation}
By integrating out the vector-like fermions, one obtains
\begin{equation}
-{\cal L}_Y = (Y_\ell)_{ij} \overline{e_{iR}} \ell_i H,
\end{equation}
and
\begin{equation}
Y_{\ell} = -\frac{y}{M_L M_E}\left( 
 \begin{array}{cc}
   y_1 y_1' \phi_1^* \phi_2^* & y_1 y_2' \phi_1^* \phi_2 \\
   y_2 y_1' \phi_1 \phi_2^* & y_2 y_2' \phi_1 \phi_2 
 \end{array}
\right).
\end{equation}
We note that
the electron is massless (at the tree level)
if only one set of the vector-like fermions are introduced as given in Eq.(\ref{Lag-f}).
Introducing one more set of the vector-like fermions, one obtains a tree-level electron mass,
though we do not write it explicitly to avoid the complication of the expression.

We suppose that the $U(1)_1 \times U(1)_2$ symmetry is broken down to $U(1)'$ by a vev
of a scalar $\phi$ whose charges are given in Table \ref{Table2}.
By redefining the normalization of the $U(1)'$ charge,
the $U(1)'$ charge for the right-handed charged lepton is 
\begin{equation}
a = \frac{ a_2+a_1}{a_2-a_1}.
\end{equation}
We note that
$U(1)_2$ ($U(1)_1$) is just broken if $a_1 = 0$ ($a_2 = 0$), and one obtains $a = 1$ ($a= -1$) trivially,
which returns to special charge assignments given in Ref.\cite{Foot:1994vd}.
We assume that the vev of $\phi$ is much larger than 
the vevs of $\phi_1$ and $\phi_2$ which break $U(1)'$,
and we ignore the contribution from the exchange of the heavier extra gauge boson in Section \ref{sec4}.

If the Lagrangian in Eq.(\ref{Lag-f}) has an exchange symmetry under $\ell_1 \leftrightarrow \ell_2$, 
$e_{1R} \leftrightarrow e_{2R}$,
(namely, $y_1 = y_2$, $y'_1= y'_2$; their phases can be different in more general exchange symmetry),
the fields $\ell_i$ and $e_{iR}$ can be written in terms the mass eigenstates $\ell_e$, $\ell_\mu$, $e_R$, $\mu_R$ 
as
\begin{equation}
 \ell_1 = \frac{\ell_e + e^{i\varphi_L} \ell_\mu}{\sqrt2}, \quad
 \ell_2 = \frac{\ell_e - e^{i\varphi_L}\ell_\mu}{\sqrt2}, \quad
 e_{1R} = \frac{e_R + e^{i\varphi_R}\mu_R}{\sqrt2},\quad
 e_{2R} = \frac{e_R - e^{i\varphi_R}\mu_R}{\sqrt2}.
\label{current-mass}
\end{equation}
We remark that the Yukawa couplings can have phases in general and 
there can be phases in the linear combinations in Eq.(\ref{current-mass})
in the basis where the electron and muon mass is real.
The gauge interaction is calculated as
\begin{align}
{\cal L} &= 
g_X ( \overline{\ell_1} \gamma_\alpha \ell_1 - \overline{\ell_2} \gamma_\alpha \ell_2) X^\alpha
+ a g_X (\overline{e_{1R}} \gamma_\alpha e_{1R} - \overline{e_{2R}} \gamma_\alpha e_{2R}) X^\alpha \\
&=
g_X ( e^{-i\varphi_L} \overline{\ell_\mu} \gamma_\alpha \ell_e + e^{i\varphi_L}\overline{\ell_e} \gamma_\alpha \ell_\mu) X^\alpha
+ a g_X ( e^{-i\varphi_R}\overline{\mu_R} \gamma_\alpha e_R + e^{i\varphi_R}\overline{e_R} \gamma_\alpha \mu_R) X^\alpha .
 \nonumber
\end{align}
By a phase redefinition, $\ell_\mu \to e^{- i \varphi_L} \ell_\mu$ and $\mu_R \to e^{- i \varphi_L} \mu_R$,
which does not change the phase of the muon mass,
we obtain 
Eq.(\ref{NGB-lagrangian})
with one physical phase $\varphi = \varphi_R - \varphi_L$.

\section{Heavy-light neutrino mixings}
\label{appendix:B}

In order to evaluate the muon decay operators which contain right-handed neutrinos,
we need to know the size of the heavy-light neutrino mixings.
Here, we list the knowledge on it.

Before we study the constraints on the mixings, we define the neutrino mixing matrix.
We work on the basis where the charged-lepton mass matrix is diagonal.
The neutrino mass term is given as
\begin{equation}
-{\cal L}_m = \frac12 \left( 
 \begin{array}{cc}
   \overline{(\nu^c)_R} & \overline{N_R}
 \end{array}
 \right)
{\cal M}
\left( 
 \begin{array}{c}
   \nu_L \\ (N^c)_L 
 \end{array}
 \right) + {\rm H.c.},
\end{equation}
where $\nu$ and $N$ are
 current-basis left- and right-handed neutrinos, 
and the $6\times 6$ neutrino mass matrix ${\cal M}$ is written as
 \begin{equation}
{\cal M} = 
\left( 
 \begin{array}{cc}
   0 & m_D \\
   m_D^T & M_N 
 \end{array}
 \right).
 \label{6x6}
\end{equation}
 The mass eigenstates $\nu', N'$ are given as
\begin{equation}
\left( 
 \begin{array}{c}
   \nu_L \\ (N^c)_L 
 \end{array}
 \right) = {\cal U} 
 \left( 
 \begin{array}{c}
   \nu'_L \\ N'_L
 \end{array}
 \right),
\end{equation}
and 
\begin{equation}
{\cal U}^T {\cal M} \, {\cal U} = {\rm diag} (M_{\cal I}) =  {\rm diag} (m_i, M_{I}) .
\end{equation}
We choose phases in $\cal U$ so that $M_{\cal I}$'s are real.
We use index 
$i$ 
 for the light neutrino mass eigenstates,
index $I$ for the ``heavy'' neutrino mass eigenstates,
and index ${\cal I}$ for both states.
For the generation index in the current basis,
we use $a$, $b$.
For convenience, we define
\begin{equation}
{\cal U} = \left( 
 \begin{array}{cc}
   U & X \\
   V & Y
 \end{array}
 \right).
\end{equation}
Namely,
\begin{align}
\nu_{aL} &= U_{a i} \nu'_{iL}+ X_{a I} N'_{IL} , \\
N^c_{aL} &= V_{a i} \nu'_{iL} + Y_{a I} N'_{IL} .
\end{align}
In the following, the mass eigenstates $\nu_i$ and $N_I$ are defined as Majorana fermions,
namely, $\nu_i \equiv \nu'_{iL} + (\nu'_{iL})^c$ and $N_I \equiv N'_{IL} + (N'_{IL})^c$.

Our concern is the constraints of the size of $V_{ai}$,
i.e., the mass eigenstate of the active neutrino in the current basis of the right-handed neutrino $N_{aL}^c$.
Because of the unitarity of the mixing matrix ${\cal U}$, we obtain
\begin{equation}
U_{ai} V^*_{bi} + X_{aI} Y^*_{bI} = 0.
\end{equation}
Therefore, let us first enumerate the constraints on $X_{aI}$ \cite{Atre:2009rg,Deppisch:2015qwa}.

\begin{enumerate}
\item
The mixings are bounded by electroweak precision data 
\begin{equation}
\sum_I |X_{eI}|^2, \sum_I |X_{\mu I} |^2 \alt 0.003,
\label{bound1}
\end{equation}
individually.
This obeys the unitarity $\sum\limits_i |U_{ai}|^2  = 1 -  \sum\limits_I |X_{aI}|^2$ 
and the universality of the four-fermion decays.
If the new muon decay operators are added, 
the bound can be modified, but 
then the contribution to the muon decay parameters from the new operators will be dominant.
If $N_I$ is lighter than the $Z$ boson, the new decay modes constrain the mixing more severely depending on their channel.

\item
The product of $|X_{eI} X_{\mu I}|$ is bounded by the $\mu \to e \gamma$ decay process as follows:
\begin{equation}
\left| \sum_I X^*_{\mu I} X_{e I}   F\left( \frac{M_I^2}{M_{W}^2} 
 \right)\right| \alt 4 \times 10^{-5},
 \label{bound2}
\end{equation}
where
\begin{equation}
 F(x) = \frac{x(1-6x+3x^2 + 2x^3 - 6x^2 \ln x)}{(1-x)^4}.
\label{tildeF}
\end{equation}

\item
For one generation ($2\times 2$ neutrino mass matrix),
the light neutrino mass in type-I seesaw is $m_\nu = m_D^2/M_N$,
and the mixing is $(X_{\alpha I})^2 = m_D^2/M_N^2 = m_\nu/M_N$, and therefore, the Mu-to-$\mubar$ transition is tiny.
For a three-generation case, there are degrees of freedom to enlarge the mixings, $X_{e I}$ and $X_{\mu I}$,
while keeping the tree-level active neutrino masses tiny.

\item
If the light--heavy neutrino mixing is enlarged,
a sizable active neutrino mass can be generated by the $Z$ boson loop diagram \cite{Pilaftsis:1991ug},
\begin{equation}
(M_\nu)_{ab}^{\rm 1-loop} \simeq \frac{\alpha_2}{4\pi \cos^2 \theta_W} \sum_I X_{a I} X_{b I} \frac{M_I^3}{M_I^2 - M_Z^2} \ln \frac{M_Z^2}{M_I^2}.
\end{equation}
The loop-induced neutrino mass can be canceled if the heavy neutrino masses are degenerate ($M_1 = - M_2$, $X_{a1} = X_{a2}$).
If the heavy neutrino masses are not degenerate and one wants to avoid unnatural cancellation between the tree-level and one-loop neutrino masses,
we need $X_{aI} \alt O(10^{-5})$ for $M_I \sim 1$ TeV.
Therefore, we usually suppose that there is a mass degeneracy in the heavy neutrino sector
to obtain a size of the mixing $X_{aI}$.

\item
The neutrinoless double beta decay ($0\nu 2\beta$) process via the heavy neutrinos 
$X_{eI}^2/M_I$,
which can be canceled for the degenerate heavy neutrino masses.
If it is not canceled, the current half-lifetime gives the bound
\begin{equation}
|X_{eI}|^2 \alt 10^{-5} \times \frac{M_I}{1\ {\rm TeV}}.
\end{equation}

\end{enumerate}

Next, let us see the direct constraints on $V_{ai}$ \cite{Barry:2013xxa,Dev:2014xea}.

In the left-right model,
the $0\nu 2\beta$ process can be induced via $W_L\,$--$\,W_R$ mixing
and the $W_R$ coupling to the right-handed electron,
 and
$U_{ei} V_{ei}^*$ is bounded as 
\begin{equation}
|U_{ei} V_{ei}^*| \alt O(10^{-4}) \times \frac{g_L}{g_R}\frac{10^{-5} }{\xi_{LR}},
\label{bound3}
\end{equation}
where $\xi_{LR}$ is a $W_L\,$--$\,W_R$ mixing,
and $g_L$ and $g_R$ are the $W_L$ and $W_R$ coupling constants.

One often considers the so-called inverse seesaw
by adding singlet fermions $N_S$.
The $9\times 9$ mass matrix for 
${\cal N} = ( \nu_L, (N^c)_L, N_S)^T$ is
\begin{equation}
{\cal M} = 
\left(
 \begin{array}{ccc}
  0 & m_D & 0 \\
  m_D^T & \mu_N & M_S \\
  0 & M_S^T & \mu_S 
 \end{array}
\right).
\end{equation}
We denote the $9\times 9$ unitary matrix 
to diagonalize 
the mass matrix ${\cal M}$ as
\begin{equation}
{\cal U} = \left( 
 \begin{array}{cc}
     U & X \\
     V & Y \\
     W & Z
 \end{array}
 \right),
 \label{9x9}
\end{equation}
where $U,V,W$ are $3\times 3$ matrices, and $X,Y, Z$ are $3\times 6$ matrices.
The light neutrino mass matrix 
is
\begin{equation}
M_\nu^{\rm light} \simeq m_D (M_S^T)^{-1} \mu_S M_S^{-1} m_D^T,
\end{equation}
for the small Majorana mass $\mu_S$.
In the left-right model, the Dirac mass $m_D$ is the naively similar size of the charged lepton masses,
which is not good to obtain sub-eV neutrino masses in the TeV-scale model.
In the inverse seesaw, the tiny neutrino masses can be explained by the smallness of $\mu_S$.
The size of $X$ is naively $m_D/M_S$, which can be sizable.
However, the size of $V$ is tiny, $\sim m_D \mu_S/M_S^2 \simeq M_\nu/m_D$, while $W$ can be as large as $X$.
{\it Therefore, the inverse seesaw is not suitable if one wants a sizable $V_{ai}$ mixing in the left-right model.}

If one wants a sizable $V_{ai}$ mixing avoiding unnatural cancellation of the tree-level and one-loop active neutrino mass, one needs a special structure of the $6\times 6$ neutrino mass matrix in Eq.({\ref{6x6}) (or a more complicated setup)
with 
the mass degeneracy in the $3\times 3$ Majorana 
neutrino mass matrix, though we do not describe the detail which is beyond the purpose of this appendix.
The structure restricts the estimation of the quantity
$J = \sum_{i,j} U_{e i} V_{\mu i}^* U_{\mu j}^* V_{e j} = \sum_{I,J} X_{e I} Y_{\mu I}^* X_{\mu J}^* Y_{e J} $, 
which affects the discussion in Section \ref{sec:left-right}.
If the sizable mixings are $X_{a1} = X_{a2}$ for $M_1 = -M_2$,
the $\mu \to e\gamma$ process bounds $X_{e1} X_{\mu 1}^*$ in Eq.(\ref{bound2}),
and therefore, the magnitude of the quantity $J$ is restricted to be less than $O(10^{-5})$.
Even if one can somehow evade the $\mu \to e\gamma$ constraint and also make $|\sum U_{\mu i} V_{e i}^*| \ll |\sum U_{e i} V_{e i}^*|$
to avoid the $0\nu 2\beta$ constraint in Eq.(\ref{bound3}), the bound in Eq.(\ref{bound1})
will restrict $|J|$ to be less than $O(10^{-3})$.

\end{document}